\documentclass[reprint,nofootinbib,amsmath,amssymb,aps,pra,onecolumn]{revtex4-2}
\usepackage[caption=false]{subfig}
\usepackage{dcolumn}
\usepackage{amsmath}
\usepackage{amsfonts}
\usepackage{amssymb}
\usepackage[colorlinks=true,citecolor=myurlcolor,linkcolor=myurlcolor,urlcolor=myurlcolor]{hyperref}
\usepackage{colortbl}\definecolor{myurlcolor}{rgb}{0,0,0.6}
\usepackage{color} 
\usepackage{graphicx}
\usepackage{hyperref}
\usepackage{adjustbox}
\usepackage{enumitem}
\usepackage{dsfont}
\usepackage{verbatim}
\usepackage{mathrsfs}
\usepackage{physics}
\usepackage{cancel}
\usepackage{mathtools}
\usepackage[normalem]{ulem}
\usepackage{cancel}

\makeatletter
\newcommand{\fmarki}{\ensuremath{\dagger}}
\newcommand{\fmarkii}{\ensuremath{\ddagger}}
\def\@fnsymbol#1{{\ifcase#1\or \fmarki\or \fmarkii \else\@ctrerr\fi}}
\makeatother   
\renewcommand{\fmarki}{\ensuremath{\dagger}}
\renewcommand{\fmarkii}{\ensuremath{\ddagger}}

\graphicspath{{./Figures/}}

\begin{document}
\title{Tunable non-Markovian dynamics in a collision model: an application to coherent transport}

\author{Simone Rijavec$^{*,}$}
\email{simone.rijavec@physics.ox.ac.uk}
\author{Giuseppe Di Pietra$^{*,}$}
\email{giuseppe.dipietra@physics.ox.ac.uk}

\affiliation{Clarendon Laboratory, University of Oxford, Parks Road, Oxford OX1 3PU, United Kingdom}

\def\thefootnote{*}\footnotetext{These authors contributed equally to this work.}

\date{\today}%

\begin{abstract}
We propose a collision model to investigate the information dynamics of a system coupled to an environment with varying degrees of non-Markovianity. We control the degree of non-Markovianity by applying a depolarising channel to a fixed and rigid reservoir of qubits.
We characterise the effect of the depolarising channel and apply the model to study the coherent transport of {information} on a chain of three interacting qubits. We show how the system-environment coupling {probability} and the degree of non-Markovianity affect the process. Interestingly, in some cases a Markovian environment is preferable to {reduce information loss and} enhance the {coherent transport}.
\end{abstract}

\maketitle

\section{Introduction}
The study of coherent energy transport in quantum systems is of great theoretical and technological importance in physics, chemistry, biology and engineering \cite{alexandrov_transport_2021, alexandrov_transport_2022, lloyd_quantum_2011,kim_quantum_2021,  muller_engineered_2012, civolani_engineering_2023}.
In photosynthetic systems, long-lived quantum coherence was observed in excitonic energy transfer \cite{engel_evidence_2007,panitchayangkoon_long-lived_2010}, although its role in the remarkable efficiency of the process has been debated \cite{wilkins_why_2015, duan_nature_2017,tao_coherent_2020}. Moreover, the role of the environment in this process has been shown to be complex \cite{lee_coherence_2007,mohseni_environment-assisted_2008,rebentrost_environment-assisted_2009, rebentrost_role_2009,strumpfer_effect_2011,rebentrost_communication_2011,butkus_vibrational_2012, huelga_vibrations_2013,tempelaar_vibrational_2014,halpin_two-dimensional_2014, duan_origin_2015, reppert_quantumness_2018,moreira_enhancing2020}.
Nevertheless, photosynthetic systems provide a promising scenario in which recently proposed witnesses of non-classicality could be tested. These witnesses suggest the possibility of indirectly inferring the non-classicality of an unknown system by looking at its capability of creating quantum coherence in a quantum probe \cite{di_pietra_temporal_2023}. 
For example, the non-classicality of a generic biopolymer could be inferred by observing the coherence of a photon emitted through the recombination of a delocalised exciton on the polymer \cite{di_pietra_temporal_2023bio}.
To investigate the possibility of this task, it is essential to understand the mechanism for coherent transport in these systems and clarify the role of the environment in the process. 

In this paper, we take an information-theoretic approach to investigate coherent transport in a simplified scenario. We propose a \textit{collision model} \cite{campbell_collision_2021,ciccarello_quantum_2022} to study the non-Markovian information dynamics of a chain of qubits interacting with the environment. 
Differently from other models in the literature \cite{ziman_diluting_2002,ziman_description_2005,ziman_all_2005,ciccarello_collision-model-based_2013,mccloskey_non-markovianity_2014,kretschmer_collision_2016,man_temperature_2018,saha_quantum_2024}, the system interacts \textit{locally} with the same reservoir qubits and we control the degree of non-Markovianity of the environment by applying a depolarising channel to the reservoir. By changing the intensity of the depolarising effect, we can control the amount of information irretrievably lost by the environment. {We quantify this effect using a measure of non-Markovianity. We also study how non-Markovianity affects the rate of information loss and find a regime where Markovianity surprisingly reduces it.}

 {We then apply} the collision model to study the coherent transport of {information, represented as an excitation,} along a chain of qubits coupled to a \textit{fixed} and \textit{rigid} reservoir of interacting qubits. By changing the degree of non-Markovianity of the reservoir, we can model different types of environments. For example, a Markovian environment could represent a thermal bath surrounding the chain, while a non-Markovian one could model other degrees of freedom of the chain that are coupled to the ones responsible for the coherent transport. {Our model is designed to study the dynamics of quantum information, not the spatial trajectory of a specific excitation, focusing specifically on the coherence of information transport along the chain.}

We show that the transfer of information to the environment plays a crucial role in the coherence of the transport. The process displays different behaviours depending on the {probability} of the system-environment {interaction}. When the system-environment coupling is strong, non-Markovianity dramatically improves the coherence of the transport. Interestingly, when the coupling is weak, {there is a regime where} a Markovian environment is preferable to enhance the {coherent transport.}

Our results provide some information-theoretic insights into how non-Markovian environments affect {coherent information} transport on a chain. Moreover, our model demonstrates how to introduce a control on the degree of non-Markovianity in collision models with a fixed environment and might be applied in other fields where transport phenomena are relevant.

\section{The model}\label{sec:model}
\subparagraph{The chain.}
Let us consider an isotropic, 1-dimensional chain of N qubits. We shall call these qubits ``system qubits" $S_n, n=1,...,N$. The chain evolves according to a \textit{Heisenberg Hamiltonian} \cite{kay_perfect_2010}:
\begin{equation}
    \hat{H}_{chain}=\frac{1}{2}\sum_{n=1}^{N} J^{n}_{chain}\left(\hat{X}_n\hat{X}_{n+1}+\hat{Y}_n\hat{Y}_{n+1}+\hat{Z}_n\hat{Z}_{n+1}\right),
    \label{eq:XXHam}
\end{equation}
with $J^n_{chain} \in \mathbb{R}\ \forall n=1, ..., N$ and $\hat{X}_n=\mathbb{I}^{\otimes n-1}\otimes \hat{\sigma}_x 
\otimes\mathbb{I}^{\otimes N-n+1}$, $\hat{Y}_n=\mathbb{I}^{\otimes n-1}\otimes \hat{\sigma}_y 
\otimes\mathbb{I}^{\otimes N-n+1}$ and $\hat{Z}_n=\mathbb{I}^{\otimes n-1}\otimes \hat{\sigma}_z 
\otimes\mathbb{I}^{\otimes N-n+1}$ the Pauli operators describing the $X$, $Y$ and $Z$ spin components of the $n$-th system qubit, respectively. By changing $J^{n}_{chain}$, this model can capture different physical systems characterised by different interaction strengths between their constituents. Moreover, the model can be easily generalised to explore other phenomena, such as anisotropies in the chain, next-nearest neighbours coupling, multiple dimensions, and effects of external fields or systems \cite{civolani_engineering_2023}.

\subparagraph{The environment.} 
We model the environment as a reservoir of $N$ interacting ``reservoir qubits" $Q_m$, $m=1, ..., N$. {To allow the information to flow in the environment as well, their interaction is described by a Heisenberg Hamiltonian:}
\begin{equation}
    \hat{H}_{res}=\frac{1}{2}\sum_{m=1}^{N} J^{m}_{res}\left(\hat{\mathcal{X}}_m\hat{\mathcal{X}}_{m+1}+\hat{\mathcal{Y}}_m\hat{\mathcal{Y}}_{m+1}+\hat{\mathcal{Z}}_m\hat{\mathcal{Z}}_{m+1}\right),
    \label{eq:XXHamres}
\end{equation}
with coupling constant $J^{m}_{res} \in \mathbb{R}\ \forall m=1, ..., N$ and $\hat{\mathcal{X}}_m$, $\hat{\mathcal{Y}}_m$, $\hat{\mathcal{Z}}_m$ the Pauli operators of the m-th reservoir qubit. 
Since we are interested in how the information dynamics of the system is affected by a loss of information, we initialize the reservoir qubits in the maximally mixed state
\begin{equation}
    \xi=\bigotimes_{m=1}^{N} \xi_m=\left[\frac{\mathds{I}}{2}\right]^{\otimes N},
    \label{eq:initialres}
\end{equation}
which represents a state containing no information.

We describe the system-environment interactions as \textit{collisions}  of each system qubit with its nearest reservoir qubit via a \textit{partial swap} \cite{ziman_diluting_2002}:
\begin{equation}
    \hat{P}_m(\eta)=\cos (\eta) \mathbb{I}+i\sin (\eta)\hat{\mathcal{S}}_m,
    \label{eq:PSWAP}
\end{equation}
where $\mathbb{I}$ is the identity operator, $\hat{\mathcal{S}}_m$ the swap operator between the m-th system and reservoir qubit, and $\eta\in[0,\pi/2]$ a parameter that regulates the {probability} of the interaction. 
Since $\hat{P}_m(\eta)$ can be written as
\begin{equation}
    \hat{P}_m(\eta)=e^{i\eta/2}e^{i\eta/2(\hat{X}_m\hat{\mathcal{X}}_{m}+ \hat{Y}_m \hat{\mathcal{Y}}_{m}+\hat{Z}_m\hat{\mathcal{Z}}_{m})},
\end{equation}
all the interactions in the model up to now are fundamentally of the same type.
{However, we prefer to maintain a continuous description in time for the chain and reservoir interactions to avoid introducing temporal order into the pairwise interactions between their qubits.}

Since all the interactions considered {so far} are unitary and the same reservoir qubits keep interacting with the chain, the model is intrinsically non-Markovian. Unlike other non-Markovian collision models where the reservoir qubits that have interacted with the system are discarded at some point \cite{ciccarello_collision-model-based_2013,mccloskey_non-markovianity_2014,kretschmer_collision_2016,man_temperature_2018,saha_quantum_2024}, we {propose to} control the \textit{degree} of non-Markovianity of the model by applying a \textit{quantum depolarising channel} to the {\textit{fixed} and \textit{rigid}} reservoir qubits after each collision. 
If $\rho_{CR}$ is the state of chain+reservoir, the depolarising channel for a single reservoir qubit can be expressed as:
\begin{equation}
    \Delta_{\Omega}^m(\rho_{CR})=\sum_{i=0}^3 \hat{K}_i^m \rho_{CR} \hat{K}_i^{m\dagger},
    \label{eq:depolarising}
\end{equation}
where $\hat{K}_i^m$ are the Kraus operators $\hat{K}_0^m \coloneqq \sqrt{1-3\Omega/4}\,\mathbb{I}_m$, $\hat{K}_1^m \coloneqq \sqrt{\Omega/4}\,\hat{\mathcal{X}}_m$, $\hat{K}_2^m \coloneqq \sqrt{\Omega/4}\,\hat{\mathcal{Y}}_m$, $\hat{K}_3^m \coloneqq \sqrt{\Omega/4}\,\hat{\mathcal{Z}}_m$, and $0\leq\Omega\leq 1$. Since the $\hat{K}_i^m$ for different $m$ commute, the order in which they are applied is irrelevant. 

\subparagraph{The protocol.}
\begin{figure}
    \includegraphics[width=0.6\columnwidth]{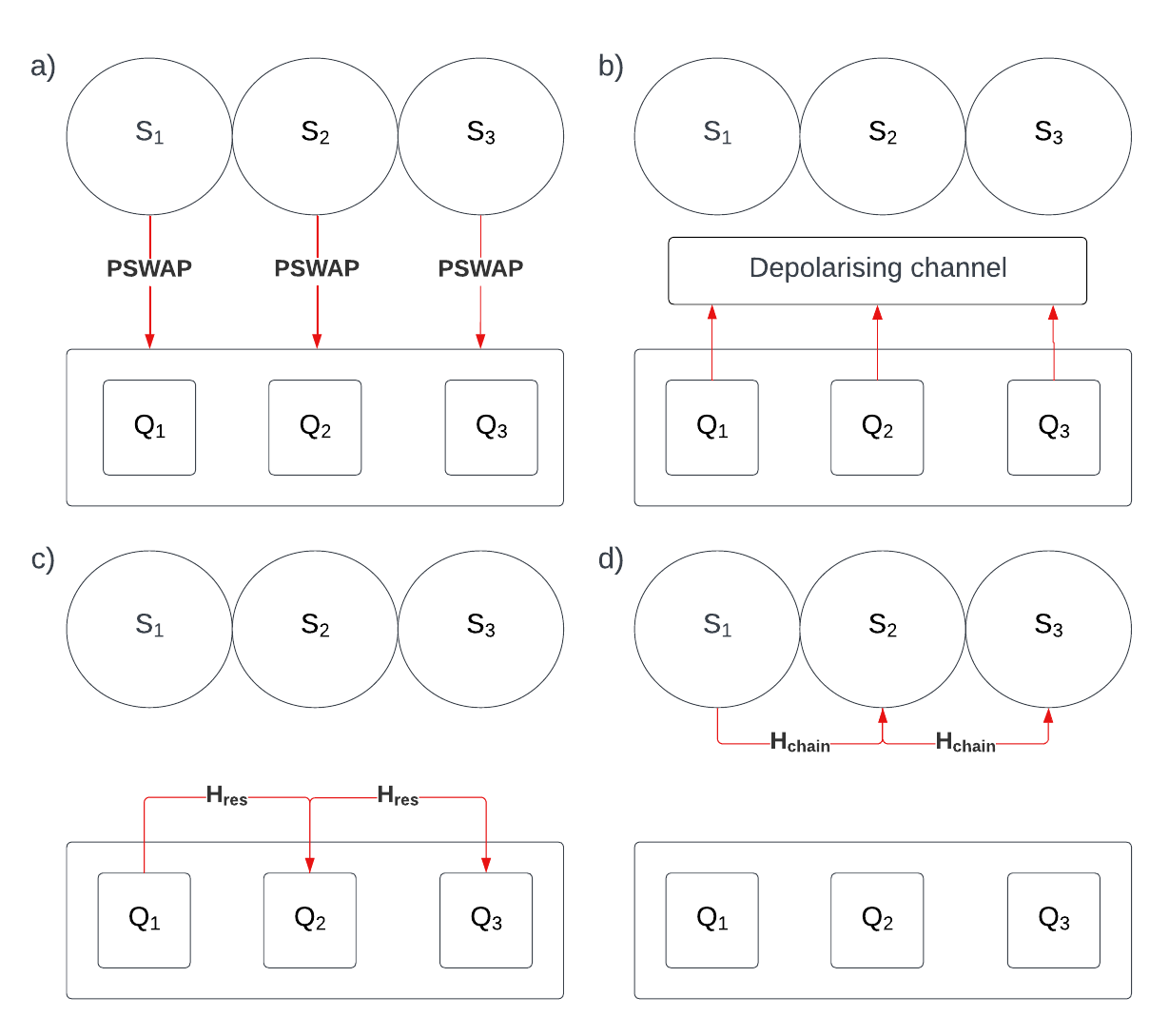}
    \caption{Four steps of the protocol with three chain qubits and three reservoir qubits. a) exchange {step}: each system qubit interacts with its nearest reservoir qubit via a partial swap; b) {depolarising} {step}: the {depolarising} channel is applied to the reservoir qubits; c) transfer via reservoir: unitary evolution of the reservoir qubits with $\hat{H}_{res}$; d) transfer via chain: unitary evolution of the system qubits with $\hat{H}_{chain}$.}
    \label{fig:mechanisms}
\end{figure}

The protocol consists of four steps (Fig. \ref{fig:mechanisms}).
\begin{enumerate}
    \item The system qubits interact with the nearest reservoir qubits via a partial swap (Eq. \eqref{eq:PSWAP}) with {probability} $\eta$ (Fig. \ref{fig:mechanisms}a). We will call this step the ``exchange {step}".
    \item The {depolarising} channel is applied to the reservoir qubits (Fig. \ref{fig:mechanisms}b). This step controls the degree of non-Markovianity of the environment. We will call this step the ``{depolarising} {step}".
    \item The reservoir qubits evolve according to the Hamiltonian of Eq. \eqref{eq:XXHamres} for a time interval $\Delta t$ (Fig. \ref{fig:mechanisms}c). We will call this step the ``transfer via reservoir".
    \item The system qubits evolve according to the Hamiltonian of Eq. \eqref{eq:XXHam} for a time interval $\Delta t$ (Fig. \ref{fig:mechanisms}d). We will call this step the ``transfer via chain".
\end{enumerate}
All the steps of the protocol can be repeated indefinitely.

{\subparagraph{Non-Markovianity.} We expect the parameter $\Omega$ to control the amount of information the environment loses. When $\Omega=0$, $\Delta_{0}^m(\rho_{CR})$ leaves $\rho_{CR}$ unchanged so that the global evolution of chain+environment is unitary. When $\Omega=1$, instead, $\Delta_{1}^m(\rho_{CR})$ re-initialises the $m$-th reservoir qubit to the maximally mixed state and, if applied to all the reservoir qubits, makes the environment Markovian. To quantify the effect of $\Omega$ on the degree of non-Markovianity, we use the measure proposed in \cite{breuer_measure_2009}. Given two initial states of the system $\rho_1(0)$ and $\rho_2(0)$, the degree of non-Markovianity under the action of a dynamical map is measured as \cite{breuer_measure_2009}:
\begin{equation}
   \mathcal{N}=\underset{\{\rho_{1}(0),\rho_{2}(0)\}}{\text{max}} \int_{\Sigma_{+}}\frac{\text{d}}{\text{d}t} D\left(\rho_1(t),\rho_2(t)\right)\text{d}t,
    \label{eq:nonM_measure1}
\end{equation}
where $\Sigma_{+}$ is the union of all subsets where $\frac{\text{d}}{\text{d}t} D\left(\rho_1(t),\rho_2(t)\right)>0$ and 
\begin{equation}
    D\left(\rho_1(t),\rho_2(t)\right)=\frac{1}{2}||\rho_1(t)-\rho_2(t)||_1 ,
\end{equation}
is the trace distance between $\rho_1(t)$ and $\rho_2(t)$.
Since in this work we consider a discrete process, we use a discretised version of Eq. \eqref{eq:nonM_measure1} \cite{breuer_measure_2009}:
\begin{equation}
    \mathcal{N}=\underset{\{\rho_{1,0},\rho_{2,0}\}}{\text{max}} \sum_{{k}\in\Sigma_+}\left(D\left(\rho_{1,{k}},\rho_{2,{k}}\right)-D\left(\rho_{1,{k}-1},\rho_{2,{k}-1}\right)\right),
    \label{eq:nonM_measure2}
\end{equation}
where the summation is taken on all values of ${k}$ such that $D\left(\rho_{1,{k}},\rho_{2,{k}}\right)-D\left(\rho_{1,{k}-1},\rho_{2,{k}-1}\right)>0$, $\rho_{1,{k}}$ is the state of the system at the end of the {$k$}-th step obtained by tracing out the environment, and the maximisation is performed over all pairs of initial states $\{\rho_{1,0},\rho_{2,0}\}$.}

{To illustrate the effect of the depolarising channel on the degree of non-Markovianity, we first restrict to a model with one system qubit and one reservoir qubit (1x1). We measure the non-Markovianity of this model by maximising Eq. \eqref{eq:nonM_measure2} over all the pairs of initial system states prepared in the following way:
\begin{equation}
    \rho_{{l},0}=
    \frac{1}{2}\begin{pmatrix}
    1+r_{{l}}\,\text{cos}\left(\theta_{{l}}\right) & r_{{l}}\,\text{sin}\left(\theta_{{l}}\right)e^{-i\Phi_{{l}}} \\
    r_{{l}}\,\text{sin}\left(\theta_{{l}}\right)e^{i\Phi_{{l}}} & 1-r_{{l}}\,\text{cos}\left(\theta_{{l}}\right) & 
    \end{pmatrix},
\end{equation}
with $r_{{l}}\in\left[0,1\right]$, $\theta_{{l}}\in\left[0,\pi\right]$, and $\phi_{{l}}\in\left[0,2\pi\right]$, for ${{l}}=1,2$.
The maximisation procedure becomes too complicated with a higher number of qubits. In this case, we use some of the insights on the states maximising Eq. \eqref{eq:nonM_measure2} drawn from the 1x1 case to choose a suitable pair of initial states. }

{
\begin{figure}
    \includegraphics[width=0.6\columnwidth]{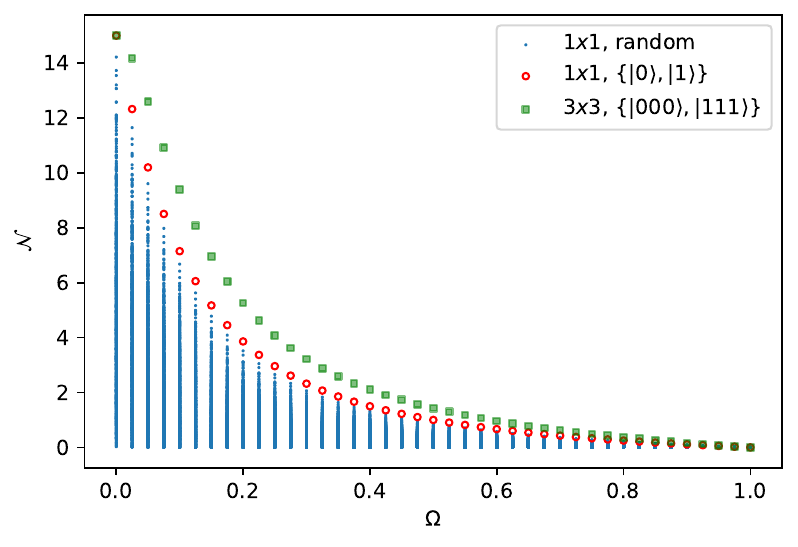}
    \caption{{Degree of non-Markovianity $\mathcal{N}$ as a function of $\Omega$ for two different models. $\mathcal{N}$ is calculated over {30} time steps with $\eta=\pi/2$. The blue circles correspond to values of $\mathcal{N}$ calculated with 1000 randomly generated pairs of initial system states in the 1x1 model. The red circles are the values of $\mathcal{N}$ obtained with the initial states $\ket{0}$ and $\ket{1}$. The green squares correspond to the values of $\mathcal{N}$ for the 3x3 model with initial states $\ket{000}$ and $\ket{111}$. {Note that the values of $\mathcal{N}$ when $\Omega\approx 0$ are bounded by the finite number of time steps of our simulations rather than by the state of the system reaching a steady state. Increasing the number of time steps would increase these values.}}}
    \label{fig:non-markovianity_measure}
\end{figure}
In Fig. \ref{fig:non-markovianity_measure}, we plot the degree of non-Markovianity $\mathcal{N}$ induced by our protocol in two models with different numbers of qubits. The circles refer to a model with one system qubit and one reservoir qubit (1x1), and the squares to the model with three system qubits and three reservoir qubits (3x3). We have calculated $\mathcal{N}$ over {30} temporal steps with $\eta=\pi/2$ for different values of $\Omega$. 
The blue circles correspond to values of $\mathcal{N}$ for 1000 randomly generated pairs of initial states of the system in the 1x1 model. The red circles correspond to the initial states $\ket{0}$ and $\ket{1}$. The graph provides strong numerical evidence that $\ket{0}$ and $\ket{1}$ maximise $\mathcal{N}$ in Eq. \eqref{eq:nonM_measure2}. During the course of the simulations, we noticed that any two antipodal states on the Bloch sphere would maximise $\mathcal{N}$. 
This observation{, and the fact that the 3x3 model consists of identical copies of the 1x1 model, lead} us to consider $\ket{000}$ and $\ket{111}$ as good candidates to maximise $\mathcal{N}$ in the 3x3 model. The green squares correspond to values of $\mathcal{N}$ for this pair of initial states and show a behaviour consistent with the 1x1 case. Due to the computational demand of simulating the 3x3 model with many different initial states, we consider only {this pair of initial states}. {More extensive simulations are required to confirm that these initial states maximise $\mathcal{N}$ as the considerations made for the 1x1 model might not apply to the 3x3 model. Therefore, the green squares can only} be interpreted as a lower bound on the non-Markovianity of the 3x3 model, although we conjecture that they maximise it.}

{Fig. \ref{fig:non-markovianity_measure} shows that the parameter $\Omega$ directly relates to the degree of non-Markovianity  $\mathcal{N}$ of the models. $\mathcal{N}$ is maximal for $\Omega=0$, that is when the evolution of system+reservoir is unitary. $\mathcal{N}$ monotonically decrease with higher values of $\Omega$ and becomes null for $\Omega=1$, i.e. when the reservoir is completely depolarised at each step. The results are similar for the 1x1 and 3x3 models and show that $\Omega$ is a good parameter to control the degree of non-Markovianity of the model.}

{\subparagraph{Information loss.} Due to the interaction with the environment, the system qubits will lose information in time. We expect the information loss rate to be related to the degree of non-Markovianity. To quantify this effect, we study the behaviour of the von Neumann entropy of a system qubit in a 1x1 model. We initialise the qubit in the state $\ket{1}\bra{1}$, corresponding to a zero-entropy state (maximal information). In Fig. \ref{fig:entropy}, we plot the von Neumann entropy $S(\rho_{C}(k))$ of the system qubit after the $k$-th step of the protocol for $\eta=0.4$ and different values of $\Omega$. We observe that the entropy tends to 1 (no information) unless $\Omega=0$. If $\Omega=0$, the system will lose some information to the environmental qubit (which is initialised in the maximally mixed state and thus has no initial information) in a periodic process, but the information will never be irretrievably lost due to the overall unitary dynamics of system+environment.
For any value of $\Omega>0$, the qubit will eventually lose all its information.
Moreover, we note that the entropy for higher values of $\Omega$ increases more slowly than for lower values of $\Omega$, meaning that information loss is slower for lower degrees of non-Markovianity. Surprisingly, Markovianity protects the qubit from information loss, although it is associated with more information loss from the environment.}

{To study how this effect depends on $\eta$, we consider the \textit{time-average} entropy of the qubit, defined as:
\begin{equation}
    \bar{S}(\eta,\Omega)=\frac{1}{K}\sum_{k=1}^{K} S(\rho_{C}(k)),
\end{equation}
where the average is taken over the first $K$ iterations of the protocol.
We plot this quantity in Fig. \ref{fig:entropy_average} for $K=30$ and different values of $\Omega$. We exclude $\Omega=0$ as, in this case, the information is never entirely lost.
Fig. \ref{fig:entropy_average} shows that, for $0.25\leq\Omega\leq1$, lower degrees of non-Markovianity protect the qubit from information loss (lower values of $\bar{S}$) if system-environment interactions are weak (lower values of $\eta$). There is an inversion in this behaviour for higher values of $\eta$, although $\bar{S}$ is higher due to the qubit losing information to the environment more quickly in this regime. This protecting mechanism of Markovianity for small $\eta$ is surprising given that Markovianity is associated with complete information loss from the environment. We might expect this effect to favour the coherent transport of information on a chain of qubits. This will be the focus of the next Section.
} 
{\begin{figure}
    \centering
    \subfloat[\label{fig:entropy} ]{\includegraphics[width=0.5\columnwidth]{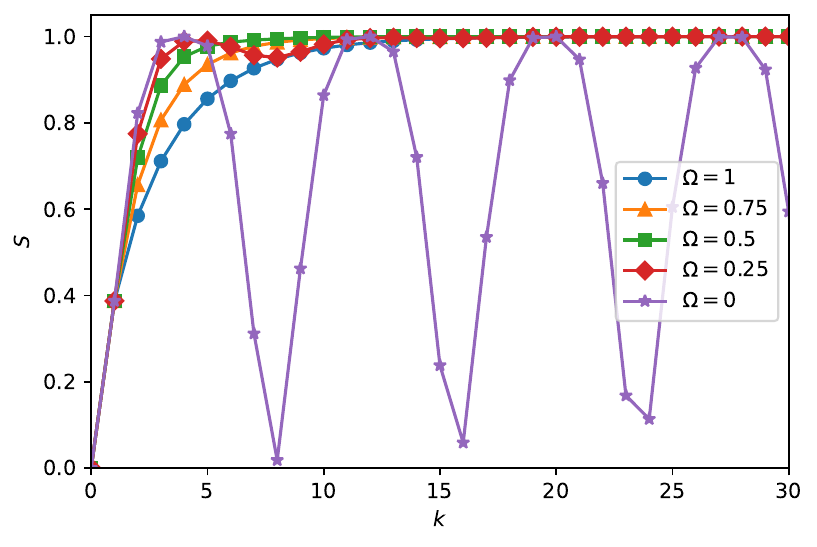}}     
    \subfloat[\label{fig:entropy_average}]{\includegraphics[width=0.5\columnwidth]{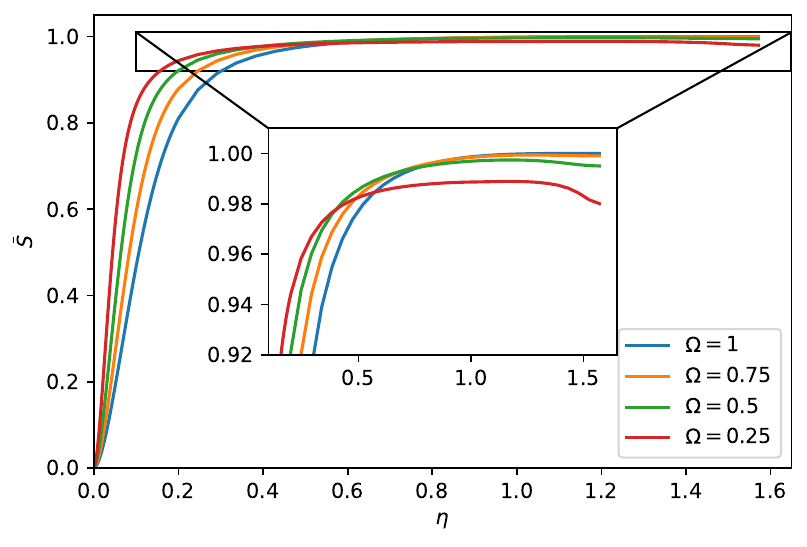}} 
    \label{fig:ent}
    \caption{{(a) Von Neumann entropy of the systems qubit $S(\rho_{C}(k))$ over multiple iterations of the protocol for $\eta=0.4$ and different values of $\Omega$.  (b) Time-averaged entropy $\bar{S}$ as a function of $\eta$ for different values of $\Omega$. The average is taken over 30 steps.}}
\end{figure}}

{\section{Application to Coherent transport}}\label{sec:application}

We apply the model described above to study the coherence of the transport of {information} along the chain. {We represent initial information as a localized excitation} 
\begin{equation}
     \rho_C(0)=\ket{100...0}\bra{100...0},
\end{equation}
in the Hilbert space $\mathscr{H}_{S_1}\otimes \dots \otimes \mathscr{H}_{S_M}$.
Applying the protocol described above, the excitation spreads through the chain and the environment. After repeating the protocol {$k$} times, we trace out the environment and extract the coherence between the first and last qubit of the chain {$\rho_C^{1,N}(k)$}, that is, the coefficient of the term $\ket{100\dots0}\bra{00\dots01}$ of the density matrix. {We take $|\rho_C^{1,N}(k)|$ as an indicator of the coherence of the transport across the chain. We choose to study this parameter as the goal of this research is to show that the excitation transport along the chain occurs \textit{coherently}, a feature that other entries of $\rho_C$ such as the population $\rho_C^{N,N}$ would not capture.}

For simplicity, we consider a model with three chain qubits. We implemented the protocol as a quantum circuit in Qiskit \cite{Qiskit} as follows. The exchange phase consists of a sequence of local unitaries $\hat{P}(\eta)$ as in Eq. \eqref{eq:PSWAP} for each of the system qubits $S_i$; the {depolarising} {step} is implemented as a depolarising noise model on the identity gate, which applies the quantum channel of Eq. \eqref{eq:depolarising}; the transfer phases are implemented as unitaries on system and reservoir with a time step $\Delta t=0.01$. We choose $J^{n}_{chain}=J_{chain}=10,\,\forall n$ and $J^{m}_{res}=J_{res}=1,\,\forall m$. This choice represents a scenario where the chain is the most efficient medium for the propagation of the excitation.

{We first apply the protocol 30 times and study how $|\rho_C^{1,3}(k)|$ varies for different values of $\Omega$ when $\eta=0.4$ and $\eta=1.2$. We select these values of $\eta$ as representative of the chain-environment ``weak" and ``strong" coupling regimes, respectively. This is because $\eta$ represents the probability of swapping the chain and reservoir qubits during the interaction described by Eq. \eqref{eq:PSWAP}: when $\eta<\frac{\pi}{4}\approx 0.8$, the probability is lower than $50\%$; when $\eta>\frac{\pi}{4}\approx 0.8$  the probability is higher than $50\%$.}
{After studying the temporal behaviour of $|\rho_C^{1,3}(k)|$ at fixed $\eta$, we investigate the effect of $\eta$ on the coherence of the transport. We do so by examining the maximum coherence $\max_k|\rho_C^{1,3}(k)|$ as a function of $\eta$ for different values $\Omega$. {We then} restrict again to the cases $\eta=0.4$ and $\eta=1.2$ to study how the ratio $J_{res}/J_{chain}$ affects $\max_k|\rho_C^{1,3}(k)|$.}
{Finally, we examine the temporal behaviour of the populations $|\rho_C^{1,1}|$ and $|\rho_C^{3,3}|$ in the same conditions, that is, the coefficients of the terms $\ket{100}\bra{100}$ and $\ket{001}\bra{001}$, respectively.}

\begin{figure*}[ht]
         \centering
         \subfloat[\label{fig:evoA} $\eta=0.4$]{\includegraphics[width=0.5\columnwidth]{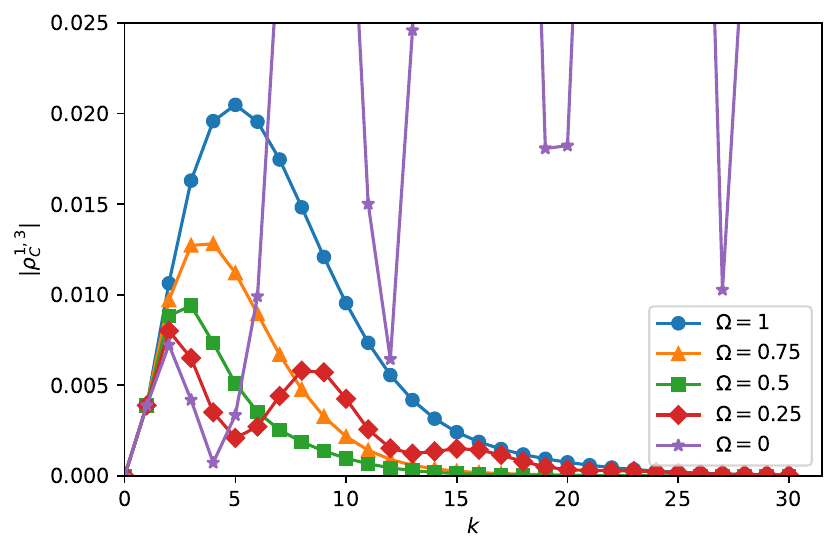}}       
          \subfloat[\label{fig:evoB} $\eta=1.2$]{\includegraphics[width=0.5\columnwidth]{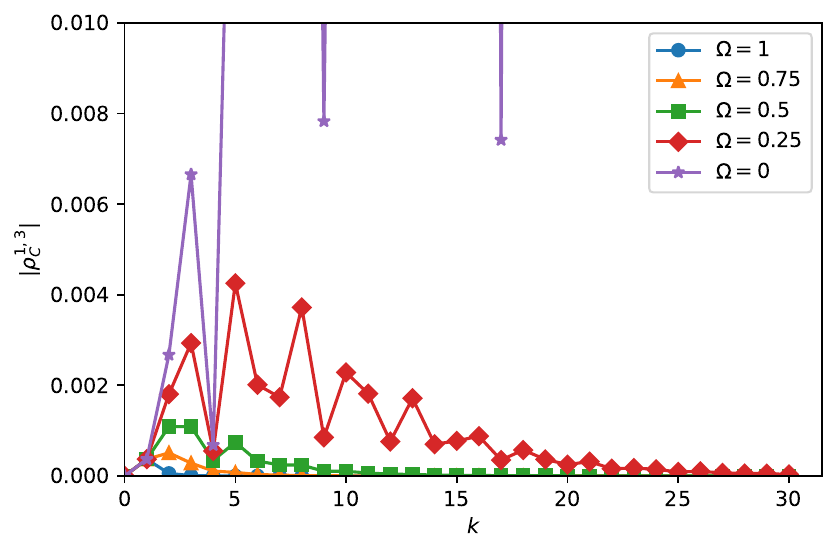}}
     \caption{Coherence element {$|\rho_C^{1,3}(k)|$} over multiple iterations of the protocol for different values of $\Omega$. The case $\Omega=0$ represents the \textit{unrealistic} scenario of a fully unitary evolution. In this case, the coherence oscillates wildly, so the vertical axis has been truncated to improve the visibility of the graphs.}
    \label{fig:evo}
\end{figure*}

{In Fig. \ref{fig:evo} we plot $|\rho_C^{1,3}(k)|$ as a function of the number of iterations of the protocol $k$ for different values of $\Omega$. We consider two cases: a) $\eta=0.4$ (weak coupling regime) in Fig. \ref{fig:evoA}, and b) $\eta=1.2$ (strong coupling regime) in Fig. \ref{fig:evoB}.}

{Before analysing the two cases separately, we notice that in both scenarios $|\rho_C^{1,3}(k)|$ is not affected by $\Omega$ for $k=1$, i.e., considering a protocol with a single iteration. This is because each protocol iteration contains a single interaction between the chain and the environment (described by $\hat{P}_m(\eta)$ in Eq. \eqref{eq:PSWAP}), so there is no further exchange of information after the application of the depolarising channel. This makes the degree of non-Markovianity of the environment irrelevant to the coherent transport on the chain.}

{In both Fig. \ref{fig:evoA} and Fig. \ref{fig:evoB}, $\Omega=0$ corresponds to a fully unitary evolution of the global system chain+environment. In this case, no information is lost, and no steady state can be reached. Such a scenario is unrealistic in most ordinary physical systems, where some information is irreparably lost over time. For this reason, we will consider only $\Omega\geq0.25$ in the following discussion. In the cases examined in Figs. \ref{fig:evoA}-\ref{fig:evoB} we see that $|\rho_C^{1,3}(k)|$ is quickly damped and reaches zero within 30 iterations of the protocol. We expect this to happen for any $\Omega\neq0$ due to the depolarising channel losing information at each protocol iteration, although the damping will take longer for values of $\Omega$ close to zero.}

{Focusing now on Fig. \ref{fig:evoA}, we observe that, in the weak coupling regime $\eta=0.4$, higher values of $\Omega$ result in greater coherence. This is {analogous to what we observed in Sec. \ref{sec:model} and it is} interesting as one may expect more information loss to be associated with decreasing amounts of coherence in the system. Instead, we observe that, for $0.25\leq\Omega\leq1$, a Markovian environment better assists coherent transport.}
{The situation is inverted in the strong coupling regime. In Fig. \ref{fig:evoB}, we observe that the coherence is immediately suppressed for higher values of $\Omega$ while it is higher and more persistent for highly \textit{non-Markovian} environments.}

\begin{figure}[ht]
    \includegraphics[width=0.6\columnwidth]{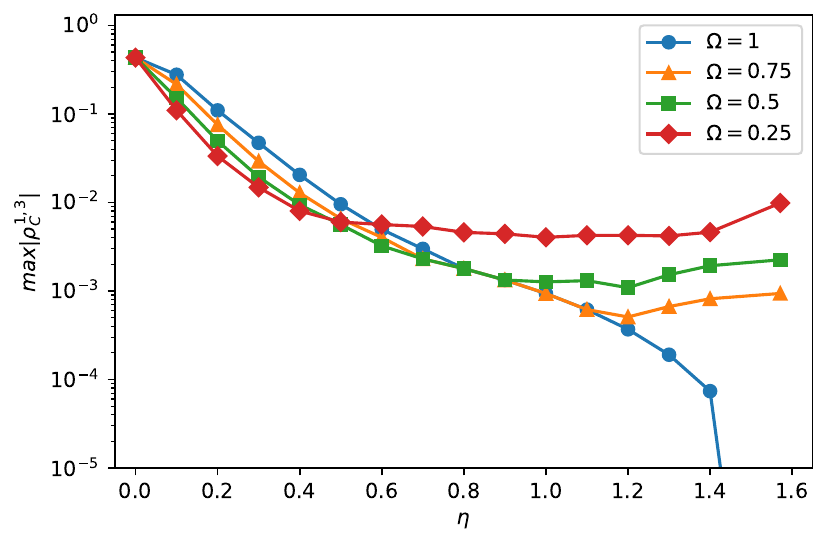}
    \caption{{Maximum coherence $\max_k|\rho_C^{1,3}(k)|$ over $k=30$ protocol iteration as a function of the system-environment coupling $\eta$, for different values of $\Omega$ when $J_{chain}=10$, $J_{res}=1$. The case $\Omega=1$ tends to zero when $\eta\rightarrow\frac{\pi}{2}$, so the y-axis has been truncated to improve the visibility of the results.}}
    \label{fig:rho_vs_eta}
\end{figure}

{So far, we have considered how $|\rho_C^{1,3}(k)|$ is affected by $\Omega$ for \textit{fixed} values of the interaction probability $\eta$, one in the weak coupling regime and the other in the strong coupling regime. To check how these results depend on the value of $\eta$, in Fig. \ref{fig:rho_vs_eta} we plot the maximum of the coherence $\max_k|\rho_C^{1,3}(k)|$ computed over 30 protocol iterations as a function of  $\eta$ for different values of $\Omega\geq0.25$.} {The results are consistent with what {was} shown above: the coherence slightly increases with the \textit{Markovianity} of the environment in the weak coupling regime, while a \textit{non-Markovian} environment is preferred in the strong coupling regime. The transition between these two regimes happens for intermediate values of the coupling probability that depend on $\Omega$.} 
{For $\eta=0$, since there is no coupling between the chain and the environment, $\max_k|\rho_C^{1,3}(k)|$ does not depend on $\Omega$. For $\eta$ close to $\frac{\pi}{2}$, i.e., with an interaction probability close to 1, the coherence goes to zero for $\Omega=1$. In this case, the coherence is immediately suppressed during the first protocol iteration because the excitation is entirely transferred to the reservoir and erased in the first two steps of the protocol (exchange and {depolarising}). For $\Omega\neq 1$, $\text{max}_k|\rho_C^{1,3}|$ is always nonzero and, after reaching a minimum for some intermediate values of $\eta$, it increases when $\eta$ tends to $\frac{\pi}{2}$, with higher values of coherence associated with lower values of $\Omega$.}

\begin{figure*}[ht]
         \centering
         \subfloat[\label{fig:vsJresA} $\eta=0.4$]{\includegraphics[width=0.5\columnwidth]{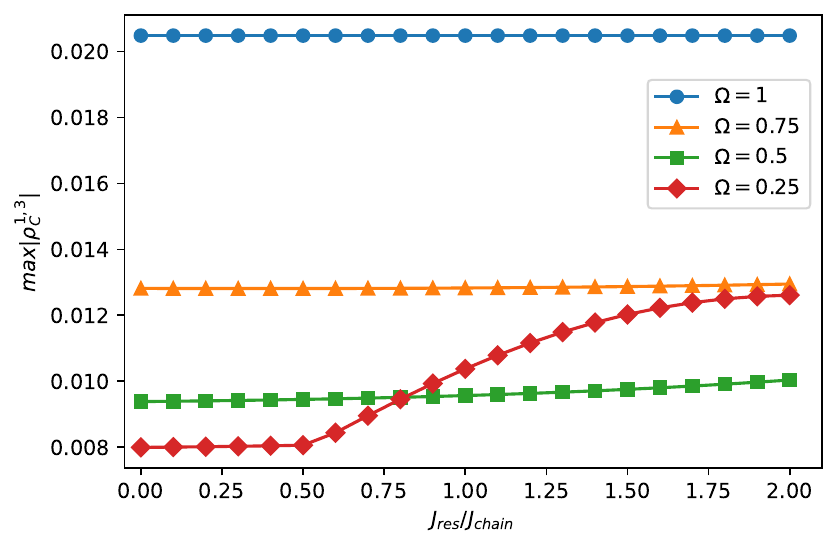}}       
          \subfloat[\label{fig:vsJresB} $\eta=1.2$]{\includegraphics[width=0.5\columnwidth]{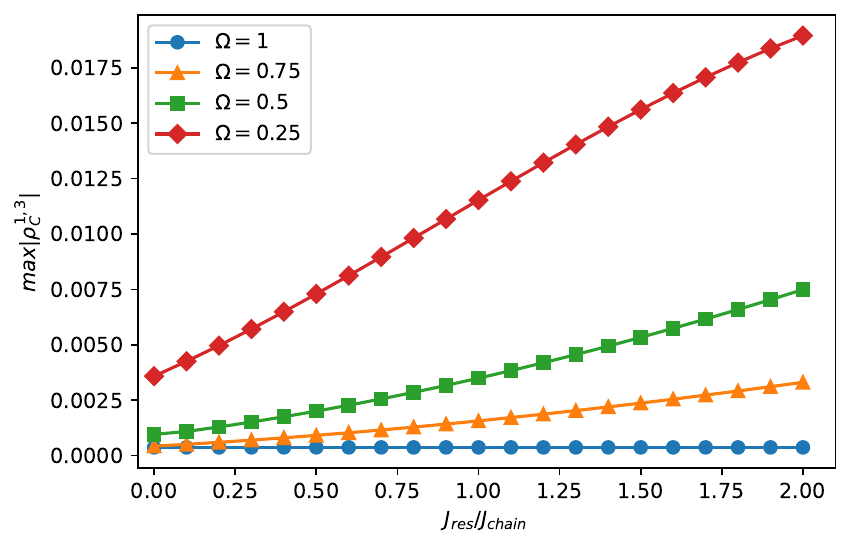}}
     \caption{{Maximum coherence $\max_k|\rho_C^{1,3}(k)|$ over $k=30$ protocol iteration as a function of the ratio between the reservoir qubits and the chain qubits couplings $J_{res}/J_{chain}$, for $J_{chain}=10$ and for different values of the degree of non-Markovianity $\Omega$.}}
    \label{fig:vsJres}
\end{figure*}

{Our results so far are based on simulations with $J_{chain}=10$ and $J_{res}=1$. As we mentioned {above}, this corresponds to a scenario where the chain is the most efficient medium for the propagation of the excitation. While this situation is the most physically realistic, in Fig. \ref{fig:vsJresA} we explore how our results depend on the ratio $J_{res}/J_{chain}$. Specifically, we plot $\max_k|\rho_C^{1,3}(k)|$ as a function of $J_{res}/J_{chain}$ in the weak ($\eta=0.4$) and strong ($\eta=1.2$) coupling regimes for $J_{chain}=10$ and different values of $\Omega\geq 0.25$. As expected, $J_{res}/J_{chain}$ does not affect $\max_k|\rho_C^{1,3}(k)|$ when $\Omega=1$ since no transfer of information along the environment is possible when all the information stored in the environment is completely lost at each protocol iteration.}

{Excluding the case $\Omega=1$, in Fig. \ref{fig:vsJresA} we observe a low sensitivity of $\max_k|\rho_C^{1,3}(k)|$ to changes in $J_{res}/J_{chain}$, with higher values of $\Omega$ initially associated to higher levels of coherence, consistently with Figs. \ref{fig:evo}-\ref{fig:rho_vs_eta}. The low sensitivity to $J_{res}/J_{chain}$ indicates that the chain is the main responsible for the coherent transport of the excitation. When $\Omega=0.25$ we observe a clear-cut change in behaviour for $J_{res}/J_{chain}=0.5$. After this point, $\max_k|\rho_C^{1,3}(k)|$ increases more rapidly with increasing values of $J_{res}/J_{chain}$, suggesting a more prominent role of the intra-environment interactions in the coherent transport.}

{The situation is different in Fig. \ref{fig:vsJresB}, where we observe that, apart from the case $\Omega=1$ discussed above, $\max_k|\rho_C^{1,3}(k)|$ consistently increases with $J_{res}/J_{chain}$, thus pointing to the active role of the environment in the coherent transport of the excitation. This happens because the environment is more likely to interact with the chain in the strong coupling regime and is thus more likely to assist in the propagation of the excitation. Accordingly, higher degrees of non-Markovianity result in higher values of coherence since the information in the environment is less likely to be lost.}

{In conclusion, when the chain is the most efficient medium for the propagation of the excitation, i.e. $J_{res}/J_{chain}\ll1$, and when $0.25\leq\Omega\leq1$, we have seen that non-Markovianity improves the coherence of the transport for strong system-environment couplings but hinders it if the coupling is weak.
Note that these two mechanisms may not be mutually exclusive. For example, in the Fenna-Matthews-Olson photosynthetic complex \cite{matthews_structure_1979}, the electronic degrees of freedom of the bacteriochlorophyll molecules (the chain in our model) can couple both to the phonons of the molecular environment (non-Markovian) and to an external thermal bath (Markovian environment). Our results might capture both of these situations, suggesting that the coexistence of a weakly coupled thermal bath and a strongly coupled non-Markovian environment might determine an optimal condition for coherent {information} transport.}

\begin{figure*}[ht]
         \centering
         \subfloat[\label{fig:pop1weak} {Population of the first system qubit $(\eta=0.4$)}]{\includegraphics[width=0.5\columnwidth]{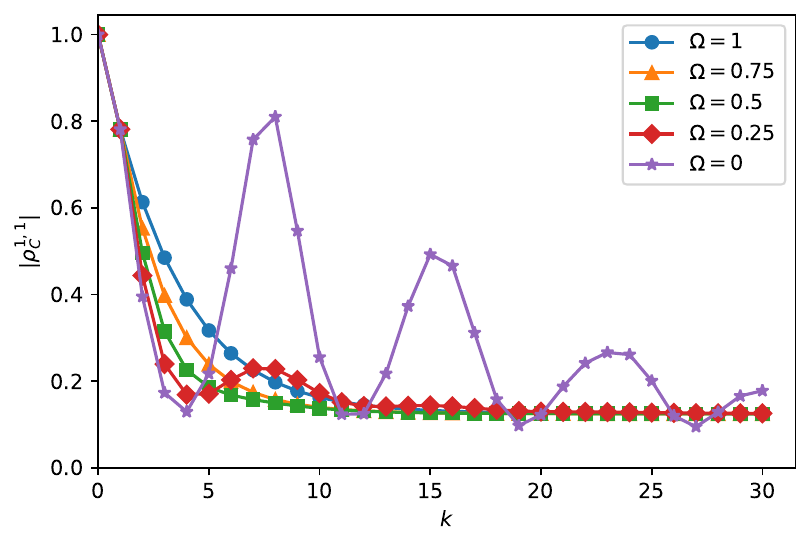}}
         \subfloat[\label{fig:pop1strong} {Population of the first system qubit $(\eta=1.2$)}]{\includegraphics[width=0.5\columnwidth]{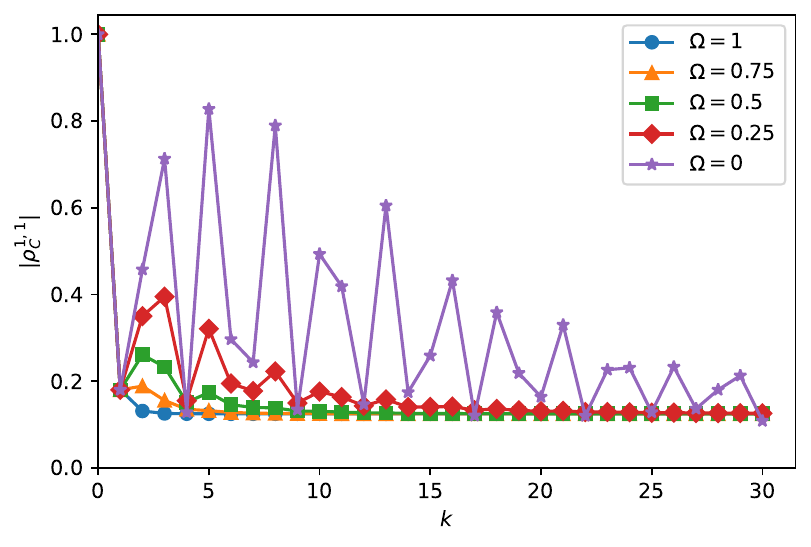}} \\
          \subfloat[\label{fig:pop3weak} {Population of the third system qubit $(\eta=0.4$)}]{\includegraphics[width=0.5\columnwidth]{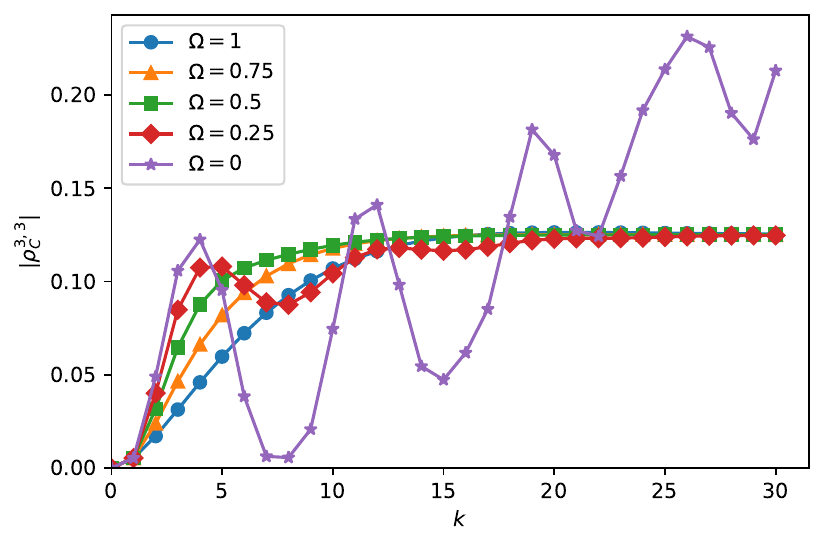}}
         \subfloat[\label{fig:pop3strong} {Population of the third system qubit $(\eta=1.2)$}]{\includegraphics[width=0.5\columnwidth]{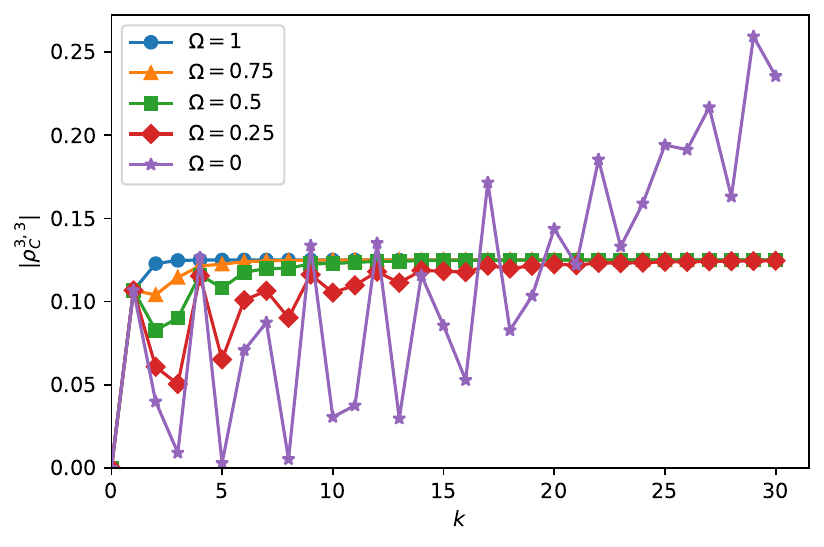}}
     \caption{{Population elements {$|\rho_C^{1,1}(k)|$} and {$|\rho_C^{3,3}(k)|$} over multiple iterations of the protocol for different values of $\Omega$.}}
    \label{fig:pop}
\end{figure*}

\begin{figure*}[ht]
         \centering
         \subfloat[\label{fig:popdiff1weak} {$|\rho_C^{1,1}(k)|$-$|\rho_C^{1,1}(k)|_{NT}$ for $\eta=0.4$}]{\includegraphics[width=0.5\columnwidth]{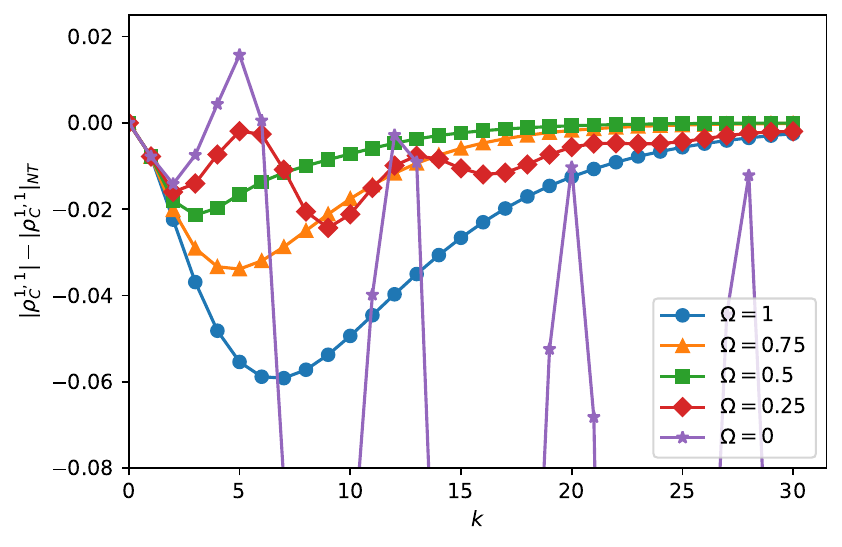}}
         \subfloat[\label{fig:popdiff1strong} {$|\rho_C^{1,1}(k)|$-$|\rho_C^{1,1}(k)|_{NT}$ for $\eta=1.2$}]{\includegraphics[width=0.5\columnwidth]{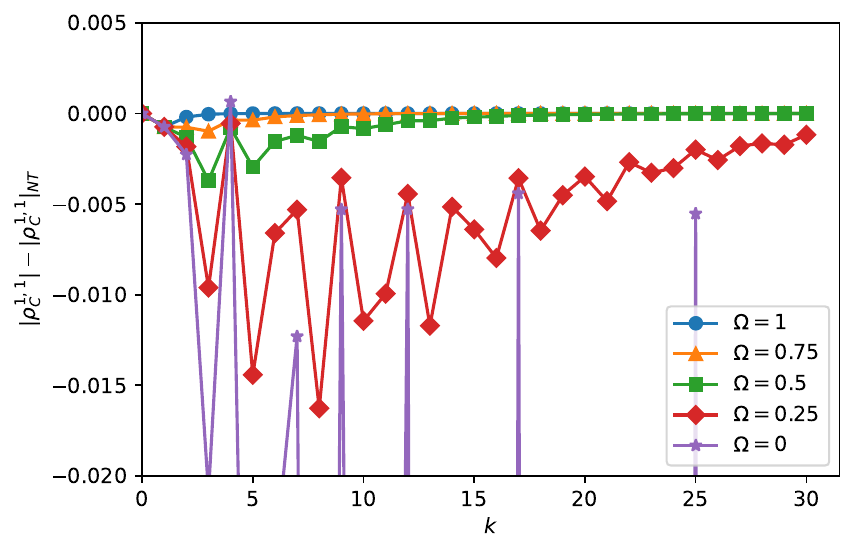}} \\
          \subfloat[\label{fig:popdiff3weak} {$|\rho_C^{3,3}(k)|$-$|\rho_C^{3,3}(k)|_{NT}$ for $\eta=0.4$}]{\includegraphics[width=0.5\columnwidth]{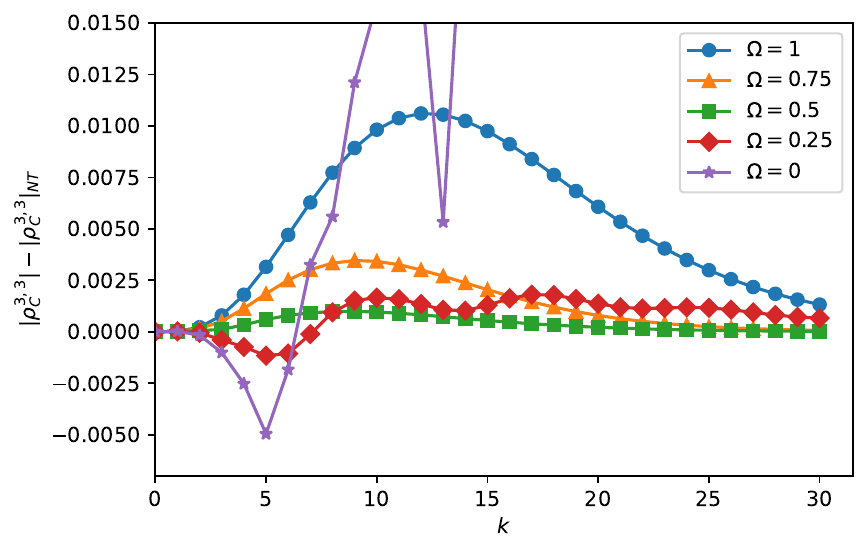}}
         \subfloat[\label{fig:popdiff3strong} {$|\rho_C^{3,3}(k)|$-$|\rho_C^{3,3}(k)|_{NT}$ for $\eta=1.2$}]{\includegraphics[width=0.5\columnwidth]{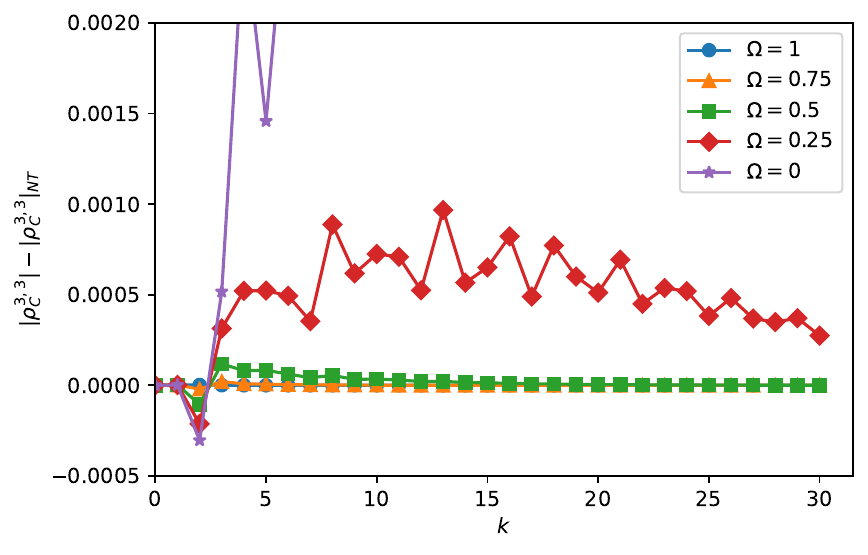}}
     \caption{{Difference between the populations $|\rho_C^{N,N}(k)|$ in the case $J_{chain}=10$, $J_{res}=1$ and the populations $|\rho_C^{N,N}(k)|_{NT}$ in the case $J_{chain}=J_{res}=0$ (no transport). The difference is plotted for the first and third qubits of the chain over multiple iterations of the protocol, for $\eta=0.4$ and $\eta=1.2$, and with different values of $\Omega$. The vertical axis has been truncated to improve the visibility of the graphs.}}
    \label{fig:popdiff}
\end{figure*}

{To complete the study, in Fig. \ref{fig:pop} we plot $|\rho_C^{1,1}(k)|$ and $|\rho_C^{3,3}(k)|$ as a function of the number of iterations of the protocol $k$ for different values of $\Omega$, once again in the weak ($\eta=0.4$, Figs. \ref{fig:pop1weak}-\ref{fig:pop3weak}) and strong ($\eta=1.2$, Figs. \ref{fig:pop1strong}-\ref{fig:pop3strong}) coupling regimes previously identified, with $J_{chain}=10$ and $J_{res}=1$. We observe that the populations tend to $1/8$ unless $\Omega=0$, which corresponds to a fully unitary process and so does not have a steady state. The steady state for $\Omega\neq0$ is the maximally mixed state $\mathds{I}^{\otimes 3}/8$ determined by the repeated application of the depolarising channel. In our model, the populations for $\Omega\neq0$ are not indicative of the location of the excitation. Instead, taken together with the coherences, they show that the information initially located on the first qubit is gradually lost. This is not a limiting aspect of our model; it just reflects its focus on the \textit{information} dynamics of the process. 
To get specific information about the location of the excitation, non-Markovianity should be implemented with a different model (see, for instance, \cite{uzdin_markovian_2018}).
Nevertheless, we expect that the excitation dynamics will also affect population changes during the initial iterations of the protocol.}

{To quantify this effect, in Fig. \ref{fig:popdiff} we plot the differences between the populations $|\rho_C^{1,1}(k)|$ and $|\rho_C^{3,3}(k)|$ of Fig. \ref{fig:pop} ($J_{chain}=10$, $J_{res}=1$), and the populations $|\rho_C^{1,1}(k)|_{NT}$ and $|\rho_C^{3,3}(k)|_{NT}$ in the case $J_{chain}=J_{res}=0$, i.e. when there is no transfer of information along the chain or the reservoir and so the change in populations is entirely due to information loss. We interpret these population differences as an indicator of the population change caused solely by excitation transport, excluding the effects of information loss.
We immediately note that---apart from the case $\Omega=0$---the values of Fig \ref{fig:popdiff} are much smaller than the corresponding values in Fig. \ref{fig:pop}, meaning that information loss is the dominant effect on population changes. Looking at Fig. \ref{fig:pop} with this in mind, we observe that higher values of $\Omega$ correspond to a slower information loss in the weak coupling regime, signalled by a slower decay in the population of the first qubit (Fig. \ref{fig:pop1weak}) and a faster increase in the population of the third qubit (Fig. \ref{fig:pop3weak}). Conversely, we see that higher values of $\Omega$ determine a quicker information loss in the strong coupling regime (Figs. \ref{fig:pop1strong}-\ref{fig:pop3strong}). This agrees with the results of Sec. \ref{sec:model}, Fig. \ref{fig:entropy}: for $0.25\leq\Omega\leq1$, higher Markovianity hampers information loss in the weak coupling regime and accelerates it in the strong coupling regime.}

{Let us now focus on the effect of the information transport on the population changes as quantified by Fig. \ref{fig:popdiff}. In Figs. \ref{fig:popdiff1weak}-\ref{fig:popdiff1strong}, negative values mean that the depopulation of the first qubit is quicker when there is transport on the chain and the reservoir ($J_{chain}\neq0, J_{res}\neq0$) compared to the case without transport ($J_{chain}=J_{res}=0$). Similarly, positive values in Figs. \ref{fig:popdiff3weak}-\ref{fig:popdiff3strong} indicate that the transport contributes to the increase in the population. Overall, Fig. \ref{fig:popdiff} points to an information transfer from the first to the third qubit before the information is lost in the long run (unless $\Omega=0$ for which there is no information loss). Moreover, we note that, for $0.25\leq\Omega\leq1$, higher Markovianity generally enhances information transfer in the weak coupling regime and lowers it in the strong coupling regime. This is in line with our results on the coherence of the transport (Fig. \ref{fig:rho_vs_eta}), meaning that this effect holds for both the amount of information transferred and its coherence.}

\section{Conclusions}
In this work, we have proposed a collision model that allows direct control of the environmental degree of non-Markovianity.
We have fixed the number and geometry of the reservoir qubits and induced information loss by applying depolarising channels on them. We have characterised the effect of the depolarising channel by measuring the degree of non-Markovianity of the model, showing how the intensity of the former directly relates to the latter. {We have also studied how the information that the system loses is related to the degree of non-Markovianity and found a regime where, surprisingly, Markovianity reduces information loss.}

We have applied our model to study the quantum coherent {information} transport on a chain of interacting qubits coupled to an environment.
Our model permits the exploration of how different regimes of non-Markovianity and system-environment coupling affect coherent transport. Studying this aspect is important, for instance, in photosynthetic complexes, where the electronic degrees of freedom responsible for the coherent transport of an exciton can couple to both Markovian (e.g. solvent) and non-Markovian (phonons) environments.
{Our results show that a non-Markovian environment aids the coherent transport only in the strong system-environment coupling regime. If the system is weakly coupled to the environment, we have observed a regime where \textit{Markovianity} enhances the coherence of the transport{, consistently with our results about information loss.}}

This work provides some information-theoretic insights into how non-Markovianity affects coherent {information} transport, avoiding the complications that arise from accurately modelling a real complex physical system. The system-independent characteristic of the model might make it suitable for applications to any quantum processes whose understanding involves the exchange of information between the system and the environment. 
The model could also be easily adapted to study more specific systems. For example, one could expand the dimensions of the chain and the environment, use more realistic Hamiltonians, or use two different environments, one Markovian and one non-Markovian. These generalisations might be relevant to inform the development of novel technological applications. The model could also be implemented on current quantum computers \cite{leontica_simulating_2021}. We leave these investigations for future work.

\section*{Acknowledgments.} 
We thank Maria Violaris, Vlatko Vedral, Chiara Marletto {and an anonymous Referee} for their sharp comments and fruitful discussions on this manuscript. S.R. thanks the Fondazione ``Angelo Della Riccia''. G.D.P. thanks the Clarendon Fund and the Oxford-Thatcher Graduate Scholarship for supporting this research.

\bibliographystyle{apsrev4-2}
\bibliography{main}

\end{document}